\documentclass[conference]{IEEEtran}
\ifCLASSINFOpdf
\fi
\usepackage{verbatim}
\usepackage[dvips]{graphics} 
\usepackage{graphicx}
\usepackage{subfigure}
\usepackage{times}
\usepackage{caption}
\usepackage{epsfig}
\usepackage[noadjust]{cite}
\usepackage{enumerate}
\usepackage{amssymb}
\usepackage{multirow}
\usepackage{latexsym}
\usepackage{color}
\usepackage{amsmath}
\usepackage[noend]{algpseudocode}
\usepackage{epstopdf}
\usepackage{tikz}
\usepackage[american]{circuitikz}
\usepackage{algorithm}
\usepackage{mathtools}
\usepackage[absolute]{textpos}

\allowdisplaybreaks
\usepackage{etoolbox}

 \makeatletter 
 \def\@eqnnum{{\normalsize \normalcolor (\theequation)}} 
  \makeatother

\begin{document} 
%\title{Joint Optimal Mode Selection and Time Allocation for RF-Powered Device-to-Device Communications}
\title{Energy-aware Mode Selection for Throughput Maximization in RF-Powered D2D Communications}
\author{\IEEEauthorblockN{Deepak Mishra$^{1}$, Swades~De$^{1}$, George~C.~Alexandropoulos$^{2}$, and Dilip Krishnaswamy$^{3}$}
\IEEEauthorblockA{$^{1}$Department of Electrical Engineering, 
Indian Institute of Technology Delhi, New Delhi 110016, India\\
$^{2}$Mathematical and Algorithmic Sciences Lab, Paris Research Center, Huawei Technologies France SASU\\
$^{3}$IBM Research Labs, Bangalore, Karnataka 560045, India
}} 

\begin{textblock}{14}(1,.1)
	\begin{center}
		Paper accepted to IEEE GLOBECOM 2017.
	\end{center}
\end{textblock}
 
\maketitle
\begin{abstract} 
Doubly-near-far problem in RF-powered networks can be mitigated by choosing appropriate device-to-device (D2D) communication mode and implementing energy-efficient information transfer (IT). In this work, we present a novel RF energy harvesting architecture where each transmitting-receiving user pair is allocated a disjoint channel for its communication which is fully powered by downlink energy transfer (ET) from hybrid access point (HAP). Considering that each user pair can select either D2D or cellular mode of communication, we propose an optimized transmission protocol controlled by the HAP that involves harvested energy-aware jointly optimal mode selection (MS) and time allocation (TA) for ET and IT to maximize the sum-throughput. Jointly global optimal solutions are derived by efficiently resolving the combinatorial issue with the help of optimal MS strategy for a given TA for ET. Closed-form expressions for the optimal TA in D2D and cellular modes are also derived to gain further analytical insights. Numerical results show that the joint optimal MS and TA, which significantly outperforms the benchmark schemes in terms of achievable RF-powered sum-throughput, is closely followed by the optimal TA scheme for D2D users. In fact, about $2/3$ fraction of the total user pairs prefer to follow the D2D mode for  efficient RF-powered IT.
\end{abstract}

\IEEEpeerreviewmaketitle
\bstctlcite{IEEEexample:BSTcontrol}
\section{Introduction}\label{sec:intro}
With the advent of 5G radio access technologies~\cite{5G_EH}, the ubiquitous deployment of low power wireless devices has led to the emergence of device-to-device (D2D) communications as a promising technology for performance enhancement by exploiting the proximity gains. Despite these merits, the underlying challenge is to provide sustainable network operation by overcoming the finite battery life bottleneck of these devices. Recently, the efficacy of energy harvesting (EH) from dedicated radio frequency (RF) energy transfer (ET) has been investigated to enable controlled energy replenishment of battery-constrained wireless devices~\cite{ComMag}. However, due to fundamental bottlenecks~\cite{ComMag}, such as low energy reception sensitivity and poor end-to-end ET efficiency, there is a need for novel RF-EH based D2D communication protocols.

In the pioneering work on wireless powered communication network (WPCN)~\cite{Th_max}, the optimal time allocation (TA) for downlink ET and uplink information transfer (IT) from multiple EH users was investigated to maximize the system throughput. It was shown that WPCN suffers from the doubly-near-far-problem which limits its practical deployment. Although optimal cooperative resource allocation strategies~\cite{ComMag,HTC} have been proposed, in this work we focus on the efficacy of D2D communications in resolving this issue by exploiting proximity. Despite the  pioneering research on WPCN~\cite{SMT,Th_max,HTC}, the investigation on EH-assisted D2D communications, where both the information source and the destination are energy constrained, is still in its infancy~\cite{CEH-D2D,EH-D2D,MTP-EH,i2RES}. In~\cite{CEH-D2D}, performance of the D2D transmission powered solely by ambient interference in cellular networks was investigated using stochastic geometry. The joint resource block and power allocation for maximizing the sum-rate of D2D links in EH-assisted D2D communication underlying downlink cellular networks was studied in~\cite{EH-D2D}. Recently, D2D communication powered by ambient interference in cellular networks for relaying machine-type communication  traffic was investigated in~\cite{MTP-EH}. An integrated information relaying and energy supply assisted RF-EH communication model was presented in~\cite{i2RES} to maximize the RF-powered throughput between two energy constrained devices. However none of the above mentioned works~\cite{CEH-D2D,EH-D2D,MTP-EH,i2RES} investigated the fundamental problem of mode selection (MS) and resource sharing between cellular and D2D users, as done in conventional D2D communications~\cite{D2D1}. In the recent work~\cite{D2D-TDD}, without considering EH, joint optimization of MS, uplink/downlink TA, and power allocation was performed to minimize the energy consumption in meeting the traffic demands of a D2D communication network. \textit{To the best of our knowledge, the harvested energy-aware joint MS and TA has not been investigated for RF-powered D2D communications. Here, different from conventional D2D systems~\cite{D2D1,D2D-TDD}, the optimal decision-making is strongly influenced by the TA for EH during downlink ET from the hybrid access point (HAP).}

\begin{figure}[!t]
\centering
{{\includegraphics[width=2.6in]{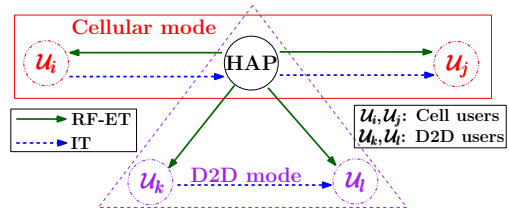} }}
\caption{\small Modes of operation in RF-powered D2D communications.}
    \label{fig:topology} 
\end{figure}

In comparison to RF-powered cellular networks, RF-EH assisted D2D communication also provides pairing gain, where, as shown in Fig. 1, each user pair can select either D2D or cellular mode of communication based on the harvested energy and radio propagation environment. D2D mode enables short-range and low-power links saving the power consumption at the transmitter, and thus assisting in overcoming the doubly-near-far problem~\cite{Th_max}. On the other hand, communicating via an energy-rich HAP in cellular mode can help in increasing the end-to-end IT range. Apart from MS, the total transmission time needs to be allocated optimally for IT and ET in D2D and cellular modes to maximize RF-powered throughput. So, we need to tackle two trade-offs: (i) choosing between D2D and cellular mode; and (ii) time sharing between ET and IT.

The key contributions of this work are four fold. (1) Novel RF-EH communication model and transmission protocol are presented to enable efficient sustainable D2D communications. (2) A joint optimization framework is proposed to maximize sum-throughput of the system by optimally selecting the transmission mode (D2D or cellular) along with TA for ET and IT. Although this is a combinatorial problem which is NP-hard in general~\cite{D2D-TDD}, we demonstrate that it can be effectively decoupled into equivalent convex problems to obtain both individual and joint global optimal MS and TA solutions. (3) Closed-form expressions are derived for: (i) optimal TA for D2D and cellular modes of communication between a single EH pair; and (ii) tight approximation for optimal TA with fixed

\noindent D2D mode for all nodes. We also provide analytical insights on the harvested energy-aware optimal MS strategy for a given TA for ET. (4) Throughput gains achieved with help of proposed optimization framework over the benchmark schemes involving fixed TA and MS are quantified by numerical simulations. These results incorporating the impact of practical RF-EH system constraints also validate the accuracy of the analysis and provide useful insights on the jointly optimal solutions.

\section{System Model}
We consider a heterogeneous small cell orthogonal frequency division multiple access (OFDMA) network with a single HAP and multiple RF-EH users that are fully-powered by  dedicated RF energy broadcast from the HAP. Without loss of generality, we focus on the RF-powered IT among $N$ RF-EH users, denoted by $\left\lbrace\mathcal{U}_1,\mathcal{U}_2,\ldots,\mathcal{U}_N\right\rbrace$, that are interested in communicating within this small cell. We assume that $\frac{N}{2}$ non-overlapping frequency channels are available to enable simultaneous communication between these possible $\frac{N}{2}$ user pairs. The transmitter and receiver in each user pair are pre-decided and allocated a disjoint frequency channel~\cite{D2D1,D2D-TDD}. Each RF-EH user is composed of a single omnidirectional antenna. To enable simultaneous reception and transmission of information at different frequencies in the uplink and downlink, the HAP is equipped with two omnidirectional antennas, one being dedicated for information reception and the other antenna is dedicated for energy or information transmission.
 
Each $\frac{N}{2}$ user pair chooses either (i) D2D mode where nodes directly communicate with each other, or (ii) cellular mode where they communicate via HAP in a two-hop decode-and-forward (DF) fashion. Both D2D and cellular modes of communication are solely powered by RF-ET from the HAP. As the cell size is very small due to low RF-ET range of the HAP~\cite{ComMag}, we have not considered the reuse mode~\cite{D2D1} to avoid strong co-channel interference. All the links are assumed to follow independent quasi-static block fading; for simplicity, we consider this block duration as $T=1$ sec. The instantaneous channel gains between the HAP and user $\mathcal{U}_i$ for downlink ET and uplink IT  are denoted by $H_{\mathcal{U}_i}$ and $\rho_{_{\mathcal{U}_i}} H_{\mathcal{U}_i}$, respectively, where $\rho_{_{\mathcal{U}_i}}$ is a positive scalar with $\rho_{_{\mathcal{U}_i}}=1$ representing the channel reciprocity case. Similarly, the channel gain between $\mathcal{U}_j$ and $\mathcal{U}_k$ is denoted by $G_{\mathcal{U}_j,\mathcal{U}_k}$. These channel gains $\{H_{\mathcal{U}_i},G_{\mathcal{U}_j,\mathcal{U}_k}\}$ can be defined as 
\begin{eqnarray}
&\hspace{-3mm}H_{\mathcal{U}_i}=\frac{p\,h_{_{0,\mathcal{U}_i}}}{d_{_{0,\mathcal{U}_i}}^n},\quad G_{\mathcal{U}_j,\mathcal{U}_k}=\frac{p\,g_{_{\mathcal{U}_j,\mathcal{U}_k}}}{d_{_{\mathcal{U}_j,\mathcal{U}_k}}^n},\;\forall i,j,k\le N, j\neq k,
\end{eqnarray}
where $p$ is path-loss coefficient; $h_{_{0,\mathcal{U}_i}}$ and $g_{_{\mathcal{U}_j,\mathcal{U}_k}}$ respectively denote the channel fading components for HAP-to-$\mathcal{U}_i$ and  $\mathcal{U}_j$-to-$\mathcal{U}_k$ links; $n$ is path-loss exponent; $d_{_{0,\mathcal{U}_i}}$ and $d_{_{\mathcal{U}_j,\mathcal{U}_k}}$ respectively denote HAP-to-$\mathcal{U}_i$ and $\mathcal{U}_j$-to-$\mathcal{U}_k$ euclidean distances.

We assume that the HAP has the full acquisition of instantaneous channel state information (CSI) of links to all users, including the CSI between all possible $\frac{N\left(N-1\right)}{2}$ user pairs. This CSI is estimated by the respective receivers and then fed back to the HAP  via the control channel~\cite{D2D1,D2D-TDD}. Although the proposed HAP-controlled optimization algorithm is based on instantaneous  CSI which incurs a lot of signalling overhead, it can serve as a benchmark for  distributed algorithms~\cite{D2D-TDD}. 

\begin{figure}[!t]
\centering
{{\includegraphics[width=3.55in]{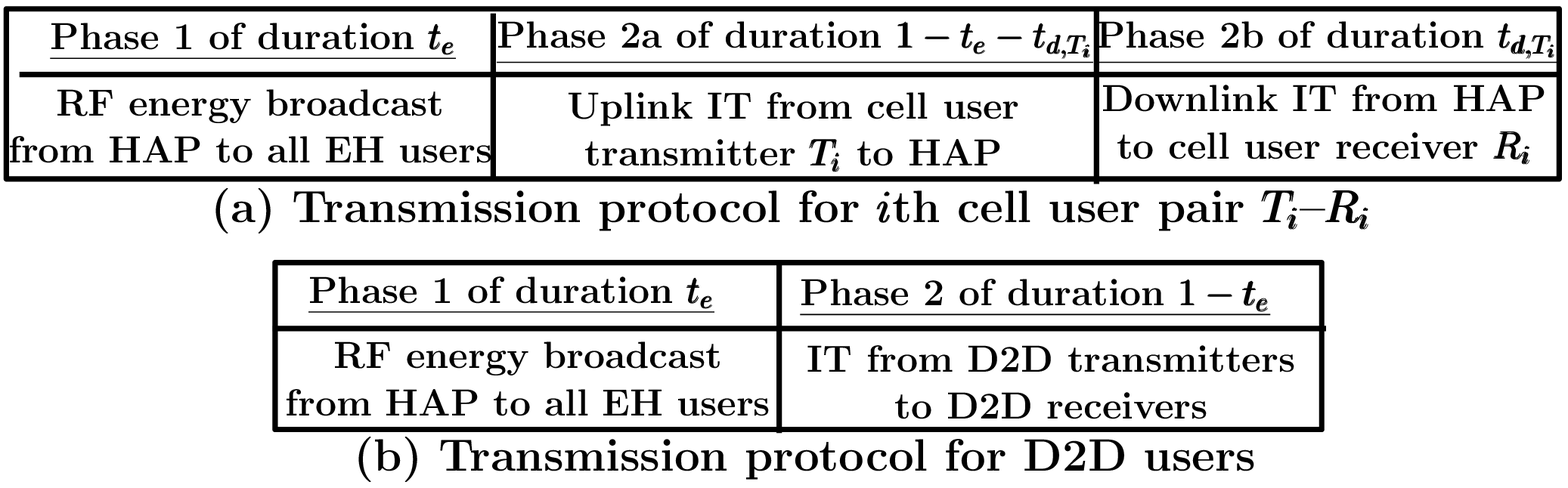} }}
\caption{\small Proposed transmission protocols for cellular and D2D modes.}
    \label{fig:proto} 
\end{figure}
As the desired frequency characteristics for efficient ET and IT are different~\cite{ComMag} and the maximum end-to-end ET efficiency is achieved over narrow-band transmission~\cite{SMT}, we consider a pessimistic scenario by ignoring EH during downlink IT from the HAP. Also, since the received uplink signal strength from energy constrained users is relatively very weak in comparison to the transmit signal strength in the downlink IT from the energy-rich HAP, we do not consider full-duplex operation at the HAP. Instead, we propose an enhanced half-duplex information relaying where orthogonal frequencies can be received and transmitted simultaneously using two antennas. 

\section{Proposed Transmission Protocol}
As shown in Fig.~\ref{fig:proto}, the proposed transmission protocols for both cellular and D2D modes of communication are divided into RF-ET of duration $t_e$  sec and  IT phase of duration $1-t_e$ sec. This downlink RF-ET based IT forms a WPCN~\cite{Th_max}. Next, we discuss the sub-operations in these two phases in detail.

\subsubsection*{Phase 1): RF-ET from the HAP to all the EH users}
During RF-ET phase, HAP broadcasts a single-tone RF energy signal $x_e$~\cite{SMT}, having zero mean and variance $P_0$, to all EH users. The energy signal $y_{e,\mathcal{U}_i}$ thus received at user $\mathcal{U}_i$ is given by
\begin{eqnarray}
&y_{e,\mathcal{U}_i}=\sqrt{H_{\mathcal{U}_i}} x_e + z_{_{\mathcal{U}_i}}, \quad\forall i\in\left\lbrace 1,2,\ldots, N\right\rbrace,
\end{eqnarray} 
where $z_{_{\mathcal{U}_i}}$ is the received Additive White Gaussian Noise (AWGN) at $\mathcal{U}_i$. Ignoring EH from noise signal $z_{_{\mathcal{U}_i}}$ due to very low energy reception sensitivity~\cite{ComMag}, the energy $E_{_{H,\mathcal{U}_i}}$ harvested at $\mathcal{U} _i$ over the RF-ET duration of $t_e$ sec is given by 
\begin{equation}\label{eqRF-EH}
E_{_{H,\mathcal{U}_i}}=\eta_{_{\mathcal{U}_i}} P_0 H_{\mathcal{U}_i} t_e, \quad\forall i\in\left\lbrace 1,2,\ldots, N\right\rbrace,
\end{equation}
where $\eta_{_{\mathcal{U}_i}}$ is RF-to-DC rectification efficiency of RF-EH unit

\noindent at $\mathcal{U}_i$ which is in general a nonlinear function of the received  RF power~\cite{NonRFH_CL}.  If $\mathcal{U}_i$ is a receiver, then harvested energy $E_{_{H,\mathcal{U}_i}}$ is stored for carrying out internal node operations. Otherwise, $E_{_{H,\mathcal{U}_i}}$ is solely used for carrying out IT from $\mathcal{U}_i$ to  HAP, if it

\noindent follows cellular mode, or to its receiving partner in D2D mode. 
\subsubsection*{Phase (2): RF-Powered IT} The operations in this phase depend on the transmission mode, i.e., cellular or D2D. 
\paragraph{Cellular Mode} The IT phase is divided into uplink and downlink subphases. By exploiting the availability of disjoint channels and two antennas at the HAP, we consider different uplink and downlink IT times, as denoted by $1-t_e-t_{d,T_i}$ and $t_{d,T_i}$, for efficient cellular mode communication between transmitter $T_i$ and receiver $R_i$ of $i$th user pair with $T_i,R_i\in\{\mathcal{U}_1,\ldots,\mathcal{U}_N\}$ and $1\le i\le N/2$. The transmit power  $P_{T_i}^C$ of $T_i$ for IT to HAP using its harvested energy $E_{_{H,T_i}}$ is 
\begin{eqnarray}\label{eqTxPs}
&P_{T_i}^C=\frac{\theta E_{_{H,T_i}}}{1-t_e-t_{d,T_i}}=\frac{\theta\eta_{_{T_i}} P_0 H_{T_i} t_e}{1-t_e-t_{d,T_i}},
\end{eqnarray}
where $\theta$ is the fraction of $E_{_{H,T_i}}$ available for IT after excluding the internal energy losses. Although we only consider the energy harvested during the current block for carrying out IT, the proposed  optimization can be easily extended to the scenario where transmit power is given as $P_{T_i}^C=\frac{\theta\big(E_{_{S,T_i}}+E_{_{H,T_i}}\big)}{1-t_e-t_{d,T_i}}$ with $E_{_{S,T_i}}$ denoting the stored energy available at $T_i$ for IT at the beginning of the current block. The information signal received at the HAP due to this uplink IT from $T_i$ is given by 
\begin{eqnarray}\label{eqISR}
& y_{u,T_i}= \sqrt{P_{T_i}^C \rho_{T_i} H_{T_i}} x_{u,T_i}+z_0,
\end{eqnarray}
where $x_{u,T_i}$ is the normalized zero mean information symbol transmitted by $T_i$ having unit variance and $z_0$ is the received AWGN at the HAP having zero mean and variance $\sigma^2$. After decoding the  message signal from $y_{u,T_i}$, the HAP forwards the decoded message signal $\widehat{x}_{u,T_i}$ to the receiving cell user $R_i$ in the second subphase of phase 2. Cellular communication between $T_i$ and $R_i$ completes upon downlink IT from the HAP to $R_i$. The information signal $y_{d,R_i}$ as received at $R_i$ is 
\begin{eqnarray}\label{eqISD}
& y_{d,R_i}=\sqrt{\phi_{_{R_i}} H_{R_i} P_0} \widehat{x}_{u,T_i} + z_{_{R_i}},
\end{eqnarray}
where $P_0$ is transmit power of HAP and $z_{_{R_i}}$ denotes zero mean AWGN having variance $\sigma^2$. $\phi_{_{R_i}}$ incorporates the difference in HAP-to-$R_i$ channel characteristics for downlink ET and IT~\cite{SMT}. 

Since the HAP acts like a DF relay~\cite{D2D1} with $1-t_e-t_{d,T_i}$ and $t_{d,T_i}$ being time allocations for uplink and downlink IT, the RF-powered throughput for cellular mode communication between $T_i$ and $R_i$, as obtained using \eqref{eqISR} and \eqref{eqISD}, is given by
\begin{align}\label{eq:cellT}
&\tau_{T_i,R_i}^C=\textstyle\min\left\lbrace\left(1-t_e-t_{d,T_i}\right)\,\log_2\left(1+\frac{\theta\eta_{_{T_i}} P_0 \rho_{_{T_i}}    H_{T_i}^2\,t_e} {\sigma^2\left( 1-t_e-t_{d,T_i}\right)}\right)\right.,\nonumber\\
&\qquad\qquad\qquad\;\;\textstyle\left.t_{d,T_i}\,\log_2\left(1+{P_0\,\phi_{_{R_i}} H_{R_i}}{\sigma^{-2}}\right)\right\rbrace.
\end{align} 
With $\tau_{T_i,R_i}^{\text{UL}}=\left(1-t_e-t_{d,T_i}\right)\,\log_2\left(1+\frac{\theta\eta_{_{T_i}} P_0 \rho_{_{T_i}}    H_{T_i}^2\,t_e}{\sigma^2\left( 1-t_e-t_{d,T_i}\right)}\right)$ and $\tau_{T_i,R_i}^{\rm{DL}}=t_{d,T_i}\,\log_2\left(1+{P_0\,\phi_{_{R_i}} H_{R_i}}{\sigma^{-2}}\right)$ respectively representing  uplink and downlink rates, throughput $\tau_{T_i,R_i}^C$ defined in \eqref{eq:cellT} can be rewritten as: $\tau_{T_i,R_i}^C=\min\left\lbrace\tau_{T_i,R_i}^{\text{UL}},\tau_{T_i,R_i}^{\rm{DL}}\right\rbrace$.

\paragraph{D2D Mode} Under this mode, $T_i$ directly communicates with $R_i$ by forming a D2D link. With IT phase of $1-t_e$ duration, the transmit power  $P_{T_i}^D$ of D2D transmitter $T_i$ is 

\noindent
\begin{eqnarray}\label{eqTxPs2}
& P_{T_i}^D=\frac{\theta E_{_{H,T_i}}}{1-t_e}=\frac{\theta\eta_{_{T_i}} P_0 H_{T_i} t_e}{1-t_e}.
\end{eqnarray} 
The corresponding information signal $y_{d2,R_i}$ received at $R_i$ is 
\begin{eqnarray}\label{eqISR2}
&y_{d2,R_i}= \sqrt{P_{T_i}^D G_{T_i,R_i}} x_{d2,T_i}+z_{_{R_i}},
\end{eqnarray}
where $x_{d2,T_i}$ is the zero mean and unit variance information symbol transmitted by  $T_i$ of $i$th user pair choosing D2D mode. 

So using \eqref{eqTxPs2} and \eqref{eqISR2}, the RF-powered throughput for the D2D mode communication between $i$th user pair is given by
\begin{align}
\tau_{T_i,R_i}^D&=\textstyle\left(1-t_e\right)\,\log_2\left(1+\frac{\theta\eta_{_{T_i}} P_0 H_{T_i} G_{T_i,R_i}\,t_e}{\sigma^2\left(1-t_e\right)}\right).
\end{align} 
With the above two throughput definitions, we investigate the joint optimal MS (between D2D and cellular) and TA (for different phases) policy for the sum-throughput $\tau_{\rm{S}}$ maximization.
 
\section{Optimal Time Allocation} 
In this section we first investigate the concavity of the RF-powered throughput in TA for both cellular and D2D modes. Then, we derive the expressions for globally optimal TA.
\subsection{Concavity of Throughput in TA}\label{sec:conc}
We first show that the throughput $\tau_{T_i,R_i}^C$ in cellular mode communication between users $T_i$ and $R_i$ is jointly concave in TAs $t_e$ and $t_{d,T_i}$ by proving the concavity of $\tau_{T_i,R_i}^{\text{UL}}$ and $\tau_{T_i,R_i}^{\text{DL}}$.  The Hessian matrix of $\tau_{T_i,R_i}^{\text{UL}}$ is given by $
\textstyle\mathbb{H}\left(\tau_{T_i,R_i}^{\text{UL}}\right)\hspace{-1mm}=\hspace{-1mm}\left[\hspace{-1.5mm}\begin{array}{ccc}
\frac{\partial^2 \tau_{T_i,R_i}^{\text{UL}}}{\partial t_e^2} & \frac{\partial^2 \tau_{T_i,R_i}^{\text{UL}}}{\partial {t_e}\partial {t_{d,T_i}}} \\
\frac{\partial^2 \tau_{T_i,R_i}^{\text{UL}}}{\partial {t_{d,T_i}}\partial {t_e}} & \frac{\partial^2 \tau_{T_i,R_i}^{\text{UL}}}{\partial {t_{d,T_i}}^2} \end{array}\hspace{-1.5mm}\right]\hspace{-1mm}=\hspace{-0.5mm}-\mathcal{Z}_i\left[\hspace{-1.5mm} \begin{array}{ccc}
1-t_{d,T_i}&  t_e\\
t_e &  \frac{t_e^2}{1-t_{d,T_i}} \end{array}\hspace{-1.5mm} \right]$
where $\mathcal{Z}_i\triangleq\frac{\left({1-t_{d,T_i}}\right)\left(\eta_{_{T_i}}\theta\rho_{_{T_i}}P_0 H_{T_i}^2 \right)^2}{\ln(2) \left(1-t_e-t_{d,T_i}\right) \left(\eta_{_{T_i}}\theta P_0 \rho_{_{T_i}}  H_{T_i}^2\, t_e+\sigma^2 \left(1-t_e-t_{d,T_i}\right)\right)^2}$. As $\mathcal{Z}_i,t_e,$ and $ t_{d,T_i}$ are positive with $t_e+t_{d,T_i}<1$, we notice that $\frac{\partial^2 \tau_{T_i,R_i}^{\text{UL}}}{\partial t_e^2},\frac{\partial^2 \tau_{T_i,R_i}^{\text{UL}}}{\partial t_{d,T_i}^2}<0$, and  the determinant of $\mathbb{H}\left(\tau_{T_i,R_i}^{\text{UL}}\right)$ is zero. This proves that $\mathbb{H}\left(\tau_{T_i,R_i}^{\text{UL}}\right)$ is negative semi-definite; hence, $\tau_{T_i,R_i}^{\text{UL}}$ is concave in $t_e$ and $t_{d,T_i}$. Also, the downlink rate $\tau_{T_i,R_i}^{\rm{DL}}$ is linear in $t_{d,T_i}$ and independent of $t_e$. Finally, with the minimum of two concave functions being concave~\cite{boyd}, the joint concavity of $\tau_{T_i,R_i}^C$ in $t_e$ and $t_{d,T_i}$ is proved.

The strict-concavity of throughput for D2D mode in $t_e$ is shown by $\frac{\partial^2 \tau_{T_i,R_i}^D}{\partial t_e^2}=\frac{-1}{{(1-t_e)\,t_e^2 \ln(2) \left(1+\frac{\sigma^2 (1-t_e)}{\eta \theta  H_{T_i} G_{T_i,R_i} P_0\, t_e}\right)^2}}<0$. 
 
\subsection{Global Optimal TA Solution}
Using \eqref{eq:cellT}, the throughput maximization problem for cellular mode communication between $T_i$ and $R_i$ can be defined as 
\begin{eqnarray*}\label{eqOPT2}
\begin{aligned} 
(\mathcal{OP}):\;&\underset{x_i,t_e,t_{d,T_i}}{\text{maximize}} \quad \tau_{T_i,R_i}^C=x_i,\quad&&\text{subject to}:\\
& (\mathrm{C1}): x_i\le \tau_{T_i,R_i}^{\text{UL}},&& (\mathrm{C2}): x_i\le \tau_{T_i,R_i}^{\rm{DL}},\\
& (\mathrm{C3}): t_e+t_{d,T_i} \le 1,&&(\mathrm{C4}): t_e,t_{d,T_i} \ge 0.
\end{aligned} 
\end{eqnarray*} 
Keeping the positivity constraint $(\mathrm{C4})$ implicit and associating Lagrange multipliers $\lambda_1,\lambda_2,\lambda_3$ with remaining constraints, the Lagrangian function for $\mathcal{OP}$ is given by $\mathcal{L}\triangleq x_i-\lambda_1\left[t+t_{d,T_i}-1\right] -\lambda_2\left[x_i-\tau_{T_i,R_i}^{\text{UL}}\right] -\lambda_3\left[x_i- \tau_{T_i,R_i}^{\rm{DL}}\right]$. The corresponding optimality (KKT) conditions~\cite{boyd} are given by constraints $(\mathrm{C1})$--$(\mathrm{C4})$, dual feasibility conditions $\lambda_1,\lambda_2,\lambda_3\ge 0$, subgradient conditions $
\frac{\partial \mathcal{L}}{\partial x_i},\frac{\partial \mathcal{L}}{\partial t_e},\frac{\partial \mathcal{L}}{\partial t_{d,T_i}}=0$, and Complementary Slackness Conditions (CSC)~\cite{boyd} $\lambda_1\left[t+t_{d,T_i}-1\right]=0,\lambda_2\left[x_i-\tau_{T_i,R_i}^{\text{UL}}\right]=0,\lambda_3\left[x_i- \tau_{T_i,R_i}^{\rm{DL}}\right]=0$. To allocate non-zero time for uplink IT, $t_e+t_{d,T_i}<1$, which results in $\lambda_1^*=0$. Using this in $\frac{\partial \mathcal{L}}{\partial x_i},\frac{\partial \mathcal{L}}{\partial t_e},\frac{\partial \mathcal{L}}{\partial t_{d,T_i}}=0$, implies that for $\tau_{T_i,R_i}^C>0$, $\lambda_1^*=0, \lambda_2^*>0,$ and $\lambda_3^*>0$. Using $\lambda_2^*>0$ and $\lambda_3^*>0$ in CSC deduces to $x_i=\tau_{T_i,R_i}^{\text{UL}}=\tau_{T_i,R_i}^{\rm{DL}}$. This implies that the optimal TA $t_{d,T_i}$ for downlink IT from the HAP to receiver $R_i$ is such that the uplink and downlink  rates become equal, i.e., $\tau_{T_i,R_i}^C=\tau_{T_i,R_i}^{\text{UL}}=\tau_{T_i,R_i}^{\rm{DL}}$. On solving this, the global optimal TA $t_{d,T_i}^*$ for downlink IT is given by
\begin{eqnarray}\label{eq:opt-td}
&\hspace{-5mm}t_{d,T_i}^*\triangleq\textstyle\left(1-t_e\right) \Bigg(1+\frac{\ln\left(\mathcal{Y}_{2,i}\right)}{\mathrm{W}_{-1}\left(-\frac{\mathcal{Y}_{1,i}  \ln\left(\mathcal{Y}_{2,i}\right)}{\mathcal{Y}_{2,i}^{\mathcal{Y}_{1,i}+1}}\right)+\mathcal{Y}_{1,i} \ln\left(\mathcal{Y}_{2,i}\right)}\Bigg)\!,
\end{eqnarray}

\noindent where $\mathcal{Y}_{1,i}=\frac{\sigma^2 (1-t_e)}{\eta_{_{T_i}} \theta P_0\,\rho_{_{T_i}} H_{T_i}^2 \, t_e}$, $\mathcal{Y}_{2,i}=1+\frac{P_0\,\phi_{_{R_i}} H_{R_i}}{\sigma^2}$, and $\mathrm{W}_{-1}\left(\cdot\right)$ is the Lambert function~\cite{Lambertfunction}. Using optimal $t_{d,T_i}^*$ along with $\tau_{T_i,R_i}^{\text{UL}}=\tau_{T_i,R_i}^{\rm{DL}}$, the optimal cellular rate $\tau_{T_i,R_i}^{C^*}\triangleq t_{d,T_i}^*\,\log_2\left(1+{P_0\,\phi_{_{R_i}} H_{R_i}}{\sigma^{-2}}\right)$ is a unimodal function of $t_e$.

Since throughput $\tau_{T_i,R_i}^D$ in D2D communication between $T_i$ and $R_i$ is strictly-concave in $t_e$, the global optimal TA $t_e$ for maximizing it, as obtained by solving $\frac{\partial\tau_{T_i,R_i}^D}{\partial t_e}\!=\!0$, is given by

\noindent 
\begin{eqnarray}\label{eq:D2D-TA}
&\hspace{-6mm}t_{e,i}^{\rm{D}^*}\!\triangleq\!\left[1\!-\!\frac{\mathcal{Y}_{3,i}\mathrm{W}_0\left(\frac{\mathcal{Y}_{3,i}-1}{\exp(1)}\right)}{\mathrm{W}_0\left(\frac{\mathcal{Y}_{3,i}-1}{\exp(1)}\right)-\mathcal{Y}_{3,i}+1}\right]^{\!-1}\hspace{-3mm}\text{ with }\mathcal{Y}_{3,i}=\frac{ P_0 H_{T_i} G_{T_i,R_i}}{\left[ \theta \eta_{_{T_i}}\right]^{-1}\sigma^2}
\end{eqnarray}
Here $\mathrm{W}_{0}\left(\cdot\right)$ is the Lambert function of principal branch~\cite{Lambertfunction}.
 
\section{Joint Mode Selection and Time Allocation} 
Using the  definition for $\tau_{T_i,R_i}^{C^*}$, the mathematical formulation of the joint optimization of MS and TA for maximizing the sum-throughput $\tau_{\rm{S}}$ of the considered system is given by
\begin{eqnarray*}\label{eqOPT1}
\begin{aligned} 
(\mathcal{JP})\!\!:\,&\underset{t_e,\{m_i\}_{i=1}^{N/2}}{\text{maximize}} &&\hspace{-3mm}\tau_{\rm{S}}\triangleq\textstyle \sum\limits_{i=1}^{N/2}\! \left(1-m_i\right)t_{d,T_i}^* \log_2\!\left(\!1+\!\frac{P_0\,\phi_{_{R_i}} H_{R_i}}{\sigma^2}\!\right)\\& &&\hspace{-5mm}\textstyle+m_i\left(1-t_e\right) \log_2\left(1+\frac{\theta\eta_{_{T_i}} P_0 H_{T_i} G_{T_i,R_i}\,t_e}{\sigma^2\left(1-t_e\right)}\right)\\
&\text{subject to:}
&&(\mathrm{C5}): m_i\in\left\lbrace0,1\right\rbrace,\quad (\mathrm{C6}): 0\le t_e\le 1,
\end{aligned} 
\end{eqnarray*}
where constraint $(\mathrm{C5})$ is defined $\forall\,i=1,2,\ldots,N/2$ and $m_i$ is the MS based binary decision variable which is defined as
\begin{equation}
m_i=\begin{cases}
0, & \tau_{_{T_i,R_i}}^{C^*} > \tau_{_{T_i,R_i}}^D \text{ (i.e., Cellular mode)}\\
1, & \tau_{_{T_i,R_i}}^{C^*}\le\tau_{_{T_i,R_i}}^D \text{ (i.e., D2D mode)}. 
\end{cases}
\end{equation}
In general $\mathcal{JP}$ is a combinatorial problem as it involves binary variable $\{m_i\}$. However, we next present a novel harvested energy-aware optimal MS strategy that resolves this issue and helps in obtaining the jointly global optimal solution of $\mathcal{JP}.$ 
\subsection{Optimal Mode Selection Strategy}\label{sec:MS}
For a given TA $t_e$, the analytical condition for throughput of RF-powered communication between $T_i$ and $R_i$ of $i$th user pair in D2D mode to be higher than that in cellular mode is: $\tau_{T_i,R_i}^D>\tau_{T_i,R_i}^{C^*}$. Using $t_{d,T_i}^*$ defined in \eqref{eq:opt-td}, this reduces to:
\begin{eqnarray}\label{eq:cond1}
&\hspace{-4mm}1+\frac{\theta\eta_{_{T_i}} P_0 H_{T_i} G_{T_i,R_i}\,t_e}{\sigma^2\left(1-t_e\right)}>\left(1+\frac{\theta\eta_{_{T_i}} P_0 \rho_{_{T_i}}    H_{T_i}^2\,t_e}{\sigma^2\left( 1-t_e\right)\left(1-\mathcal{F}_i\right)}\right)^{1-\mathcal{F}_i},
\end{eqnarray}
where $\mathcal{F}_i\triangleq \frac{t_{d,T_i}^*}{\left(1-t_e\right)}\le 1$.
We note that $\frac{t_e}{1-t_e}$ and $\mathcal{F}_i$ are strictly

\noindent increasing functions of $t_e,$ where the latter holds because $
\frac{\partial\mathcal{F}_i}{\partial t_e}=\textstyle\frac{-\left(1-t_e-t_{d,T_i}^*\right)\sigma^2\left[\mathrm{W}_{-1}\left(-\mathcal{Y}_{2,i}^{-\mathcal{Y}_{1,i}-1}\ln\left(\mathcal{Y}_{2,i}^{\mathcal{Y}_{1,i}}\right)\right)+1\right]^{-1}}{\eta_{_{T_i}} \theta P_0 \rho_{_{T_i}}  H_{T_i}^2\,t_e^2\left(1-t_e\right)\mathcal{Y}_{1,i}}>0$. This result is obtained by knowing ${\mathrm{W}_{-1}}\left(x\right)+1<0, \forall x\in\left[\frac{-1}{\exp(1)},0\right]$ and $\mathcal{Y}_{2,i}>1$ along with $\frac{-1}{\exp(1)}\!\le\!\frac{-\ln\left(\mathcal{Y}_{2,i}^{\mathcal{Y}_{1,i}}\right)}{\mathcal{Y}_{2,i}^{\mathcal{Y}_{1,i}}}\!<\!0$.

Now if $G_{T_i,R_i}\ge\rho_{_{T_i}}H_{T_i}$, implying that direct link quality is better than the uplink quality, then from \eqref{eq:cond1} we note that $\tau_{T_i,R_i}^D\ge\tau_{T_i,R_i}^{C^*}$ or $m_i=1$ for any feasible $t_e$ because D2D mode is more energy-efficient and provides higher throughput for the same amount of harvested energy. Whereas if $G_{T_i,R_i}<\rho_{_{T_i}}H_{T_i}$ and $\exists\,t_{e,i}^{\text{th}}\triangleq \left\lbrace t_e\mathrel{}\middle|\mathrel{}\tau_{T_i,R_i}^D=\tau_{T_i,R_i}^{C^*}\right\rbrace$, then it can be shown that for $t_e>t_{e,i}^{\text{th}}$, D2D mode is preferred by the $i$th user pair over the cellular mode, and vice-versa. This holds because if $\exists\, t_{e,i}^{\text{th}}\in\left(0,1\right)$, then $\forall t_e\in\left(t_{e,i}^{\text{th}},1\right)$, $\tau_{T_i,R_i}^D>\tau_{T_i,R_i}^{C^*}$ due to strictly decreasing nature of $\tau_{T_i,R_i}^{C^*}$ in $\mathcal{F}_i$, the rate of increase of $\tau_{T_i,R_i}^{C^*}$ is lower than that of $\tau_{T_i,R_i}^D,\forall t_e>t_{e,i}^{\text{th}}.$ The value of this threshold $t_{e,i}^{\text{th}}$ implies that even when uplink quality is better than direct link  between $T_i$ and $R_i$, D2D is better mode than cellular for $t_e> t_{e,i}^{\text{th}}$ because the resulting harvested energy makes it more spectrally-efficient as compared to the cellular mode involving redundant transmission during downlink IT.

\subsection{Jointly Global Optimization Algorithm}
Using thresholds $\{t_{e,i}^{\text{th}}\}_{_{i=1}}^{^{N/2}}$, the optimal MS for each EH user pair can be decided based on TA $t_e$ for ET. So, we divide the feasible range for $t_e$ into $\frac{N}{2}+1$ subranges based on $\frac{N}{2}$ thresholds $\{t_{e,i}^{\text{th}}\}$. For each of these subranges, the optimal MS policy $\{m_i^*\}_{_{i=1}}^{^{N/2}}$ can be obtained using the discussion given in Section~\ref{sec:MS}. Further, using the concavity of $\{\tau_{T_i,R_i}^D\}_{_{i=1}}^{^{N/2}}$ and $\{\tau_{T_i,R_i}^{C^*}\}_{_{i=1}}^{^{N/2}}$ in TA, as proved in Section~\ref{sec:conc}, we note that the sum-throughput $\tau_{\rm{S}}$ for optimal MS $\{m_i^*\}$ is concave in TA $t_e$ because it is a sum of $\frac{N}{2}$ concave functions comprising of either $\{\tau_{T_i,R_i}^D\}$ or $\{\tau_{T_i,R_i}^{C^*}\}$. Hence, the global optimal MS and TA can be obtained by selecting the subrange that gives the maximum sum throughput. We have summarized these steps of the proposed joint global optimization strategy in Algorithm~\ref{Algo:Opt}.

\begin{algorithm}[!h]
{\small
\caption{Joint global optimal MS and TA to maximize $\tau_{\rm{S}}$}\label{Algo:Opt}
\begin{algorithmic}[1]
\Require Set $\mathbb{S}=\left\lbrace \left(T_i,\,R_i\right)\mathrel{}\middle|\mathrel{} T_i, R_i\in\{\mathcal{U}_1,\ldots \mathcal{U}_{N}\}\wedge 1\le i\le\frac{N}{2}\right\rbrace$; channel and system parameters $H_{T_i}, H_{R_i}, G_{T_i,R_i},\eta_{_{T_i}},\rho_{_{T_i}},\phi_{_{R_i}}$ for each user pair in $\mathbb{S}$, along with $\theta,P_0,\sigma^2$; and tolerances $\epsilon,\xi$ 
\Ensure Optimal MS $\{m_i^*\}$ and TA $t_e^*,\{t_{d,T_i}^*\}$ along with $\tau_{\rm{S}}^*$ 
\State Initialize $\mathbb{C}=\emptyset$ and define a very small positive quantity $\epsilon\approx 0$
\For{$i \in \{1,2,\dots,\frac{N}{2}\}$}
        \If {$\tau_{T_i,R_i}^D>\tau_{T_i,R_i}^{C^*}$ for both $t_e=\epsilon$ and $t_e=1-\epsilon$}
        \State Set $t_{e,i}^{\text{th}}=0$\Comment{Represents D2D mode}
        \ElsIf {$\tau_{T_i,R_i}^D<\tau_{T_i,R_i}^{C^*}$ for both $t_e=\epsilon$ and $t_e=1-\epsilon$}
        \State Set $t_{e,i}^{\text{th}}=1,\;\mathbb{C}=\mathbb{C}\cup i$\Comment{Represents Cellular mode}
        \Else
        \State Set $t_{e,i}^{\text{th}}=\left\lbrace t_e\mathrel{}\middle|\mathrel{} \tau_{T_i,R_i}^D=\tau_{T_i,R_i}^{C^*}\right\rbrace,\;\mathbb{C}=\mathbb{C}\cup i$
         \EndIf
      \EndFor
      \State Sort $t_{e,i}^{\text{th}}$ in descending to store values in $\mathcal{V}$ and indexes in $\mathcal{I}$
      \State Set $k_1\!=\!\!\underset{1\le i\le N/2}{\rm{argmin}}\!\left[\!\eta_{_{T_i}} H_{T_i} G_{T_i,R_i}\!\right]$,\,  $k_2\!=\!\!\underset{1\le i\le N/2}{\rm{argmax}}\!\left[\!\eta_{_{T_i}} H_{T_i} G_{T_i,R_i}\!\right]$
      \State Set $\tau_1\left(t_e\right)=\textstyle \sum_{j=1}^{N/2} \left(1-t_e\right)\,\log_2\Big(1+\frac{\theta\eta_{_{T_j}} P_0 H_{T_j} G_{T_j,R_j}\,t_e}{\sigma^2\left(1-t_e\right)}\Big)$
      \If {$\mid\mathbb{C}\mid=\emptyset$}\Comment{Represents all Nodes in D2D mode}
      \State Set $\text{lb}_1=t_{e,k_2}^{\rm{D}^*}$ and $\text{ub}_1=t_{e,k_1}^{\rm{D}^*}$ by using \eqref{eq:D2D-TA}
      \ElsIf {$\mid\mathbb{C}\mid=\frac{N}{2}$}\Comment{Represents all Nodes in cellular mode}
        \State Set $\text{lb}_1=0$ and $\text{ub}_1=t_{e,k_1}^{\rm{D}^*}$ 
        \State Set $\tau_1\left(t_e\right)=\textstyle \sum_{j=1}^{N/2}t_{d,T_j}^*\,\log_2\big(1+{P_0\,\phi_{_{R_j}} H_{R_j}}{\sigma^{-2}}\big)$
        \Else 
        \State Set $\text{lb}_1=\mathcal{V}_1$ and $\text{ub}_1=t_{e,k_1}^{\rm{D}^*}$ 
      
        \For{$i \in \{1,2,\dots,\mid\mathbb{C}\mid\}$}
        \State Set ${\tau_{i+1}} \left(t_e\right)={\tau_i} \left(t_e\right)+\tau_{T_{\mathcal{I}_i},R_{\mathcal{I}_i}}^{C^*}-\tau_{T_{\mathcal{I}_i},R_{\mathcal{I}_i}}^D$        \If{$i<\mid\mathbb{C}\mid$}
        \State Set $\text{lb}_{i+1}=\mathcal{V}_{i+1}$ and $\text{ub}_{i+1}=\mathcal{V}_{i}$
        \Else
        \State Set $\text{lb}_{i+1}=0$ and $\text{lb}_{i+1}=\mathcal{V}_{i}$
         \EndIf
      \EndFor
         \EndIf
 \For{$i \in \left\{1,2,\dots, \mid\mathbb{C}\mid+1 \right\}$}
	\State Set $t_l=\text{lb}_i$ and $t_u=\text{ub}_i$\label{step:st}
	\State Set $t_p=t_u-0.618\left(t_u-t_l\right),t_q=t_l+0.618\left(t_u-t_l\right)$
	\While{$t_u-t_i>\xi$}
\If{${\tau_{i}}\left(t_p\right)\geq {\tau_{i}}\left(t_q\right)$}
\State Set $t_u=t_q$, $t_q=t_p,t_p=t_u-0.618\left(t_u-t_l\right)$
\Else
\State Set $t_l=t_p$, $t_p=t_q,t_q=t_l+0.618\left(t_u-t_l\right)$
\EndIf
\EndWhile
\State Set $t_i^*=\frac{t_u+t_l}{2}$ and $\tau_{{\rm{S}},i}^*=\tau_i\left(t_i^*\right)$\label{step:en}
\If{$\left(\mid\mathbb{C}\mid=\emptyset\right)\vee\left(\mid\mathbb{C}\mid=\frac{N}{2}\right)$}
\State \textbf{break}
\EndIf
      \EndFor
     
      \State Set $\text{opt}=\underset{1\le i\le\mid\mathbb{C}\mid+1}{\operatornamewithlimits{argmax}}
\;\tau_{{\rm{S}},i}^*$,\quad $\tau_{\rm{S}}^*=\tau_{{\rm{S}},\text{opt}}^*$,\; and \; $t_e^*=t_{\text{opt}}^*$ 
\If{$\left(\mid\mathbb{C}\mid=\emptyset\right)$}
 \State Set $m_i^*=1,\;t_{d,T_i}^*=0,\;\forall i=1,2,\ldots,\frac{N}{2}$
 \ElsIf{$\left(\mid\mathbb{C}\mid=\frac{N}{2}\right)$}
 \State $\forall i, m_i^*\hspace{-0.5mm}=\hspace{-0.5mm}0$ and $t_{d,T_i}^*$ is obtained using $t_e\hspace{-0.75mm}=\hspace{-0.75mm}t_e^*$ in \eqref{eq:opt-td}
 \Else 
 \State $\forall i\!=\!1,2,\ldots,\frac{N}{2},$ set $m_i^*\!=\!\begin{cases}
1,  & \text{$i\!=\!\mathcal{I}_{\text{opt}+1},\mathcal{I}_{\text{opt}+2},\ldots,\mathcal{I}_{\mid\mathbb{C}\mid}$}\\
0,  & \text{otherwise},
\end{cases}$
 \State $t_{d,T_i}^*=\begin{cases}
\text{obtained by substituting $t_e^*$ in \eqref{eq:opt-td}},   & \text{$m_i^*=0$}\\
0, \qquad\qquad\text{ (i.e., D2D mode,) }& \text{otherwise},
\end{cases}$
\EndIf        
\end{algorithmic}
}
\end{algorithm} 

In Algorithm~\ref{Algo:Opt} after dividing the feasible $t_e$ range, optimal TA $t_e^*$ for each of the possible $\frac{N}{2}+1$ MS scenarios is obtained using the Golden Section (GS) based one dimensional (1D) search  with acceptable tolerance $\xi\ll 1$ (implemented in steps~\ref{step:st} to \ref{step:en}). The upper and lower bounds for each search are either based on the thresholds $\{t_{e,i}^{\text{th}}\}$ or on $\{t_{e,i}^{\rm{D}^*}\}$, as defined in \eqref{eq:D2D-TA} for pair $i$ having worst and best average link qualities, respectively. So, we conclude that Algorithm~\ref{Algo:Opt} returns the global optimal MS and TA along with maximum $\tau_{\rm{S}}$ after running GS-based 1D-search for at most $\frac{N}{2}+1$ times.

\section{Numerical Results and Discussion}
For generating numerical results we consider that $N$ nodes are deployed randomly following Poisson Point Process (PPP) over a square field of area $L\times L$ m$^2$ with the HAP positioned at the center. This field size $L$ ensures that the average received power for a given path-loss exponent $n$ and average fading parameter $H_{T_i}$ is higher than the minimum received power sensitivity of $-20$ dBm for practical RF-EH circuits~\cite{i2RES}. So, as $n$ is varied from $2$ to $5$, maximum field size $L$ decreases from $23.4$ m to $4.4$ m. For user pairing we have considered that node $i$ pairs with node $N-i+1,\forall i=\{1,2,\ldots, N/2\}$. The graphs in this section are obtained by plotting the average results for multiple random channel realizations and multiple random node deployments with unit average channel fading components $\{H_{T_i},H_{R_i},G_{T_i,R_i}\}$ and path-loss coefficient $p$. We have assumed $P_0\!=\!4$ W, $\sigma^2\!=\!-100$ dBm, $\theta\!=\!0.8$, $\rho_{_{T_i}}\!=\!\phi_{_{R_i}}\!=\!1,\eta_{_{T_i}}\!=\!0.5,\,\forall\, i,$  with tolerances $\xi\!=\!10^{-3}$ and $\epsilon\!=\!10^{-6}$. 
 
\begin{figure}[!t]
\centering
{{\includegraphics[width=2.7in]{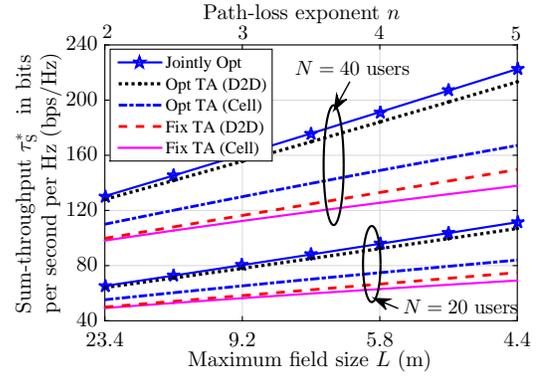} }}
\caption{\small Throughput performance comparison of different schemes.}
    \label{fig:sumrate} 
\end{figure} 
 
\begin{figure}[!t]
\centering
{{\includegraphics[width=3.4in]{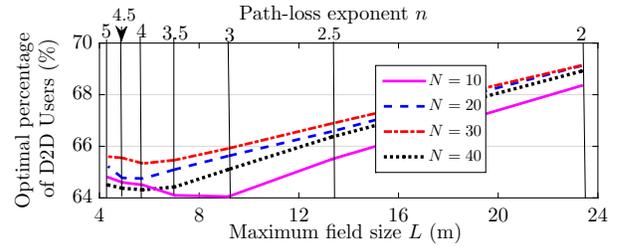} }}
\caption{\small Variation of optimal MS (D2D versus cellular) with $L,n$.}
    \label{fig:OD2D} 
\end{figure} 

\begin{figure}[!t]
\centering
{{\includegraphics[width=3.2in]{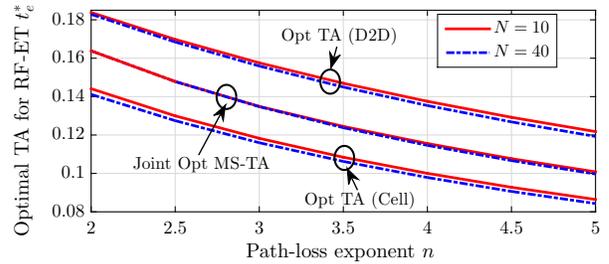} }}
\caption{\small Variation of optimal TA $t_e^*$ for ET in different modes.}
    \label{fig:OTA} 
\end{figure}
Firstly, we investigate the throughput gain achieved with the help of proposed joint MS and TA optimization over the benchmark schemes that include fixed communication mode (all nodes are either in cellular or D2D mode) and uniform TA for each phase where $t_e=t_{d,T_i}=\frac{1}{3}$ for cellular mode and $t_e=\frac{1}{2}$ for D2D mode. Results in Fig.~\ref{fig:sumrate} show that the joint optimization scheme provides significant gains over the fixed TA and MS schemes. These gains which scale with increased system size $N$, get enhanced with diminishing field size $L$ due to increased path-loss exponent $n$ for both $20$ and $40$ user systems. Furthermore, the throughput performance of D2D mode for both optimal TA and fixed TA is much better than that for the corresponding cellular mode communication. In fact, the performance of optimal TA with all nodes selecting D2D mode is very close to that of joint optimal strategy. The reason for this can be observed from the results plotted in Figs.~\ref{fig:OD2D} and \ref{fig:OTA}. From Fig.~\ref{fig:OD2D} we observe that irrespective of field size $L$, about $2/3$ fraction of the total users prefer D2D mode. As the D2D mode involves direct IT, it is more spectrally-efficient and can allocate higher $t_e^*$ for ET as shown in Fig.~\ref{fig:OTA}. Since joint MS and TA involves both D2D and cellular modes, its $t_e^*$ lies between that of all-D2D and all-cellular scenarios.

\begin{figure}[!t]
  \centering 
   \subfigure[Average uplink and downlink IT time]{{\includegraphics[width=2.6in]{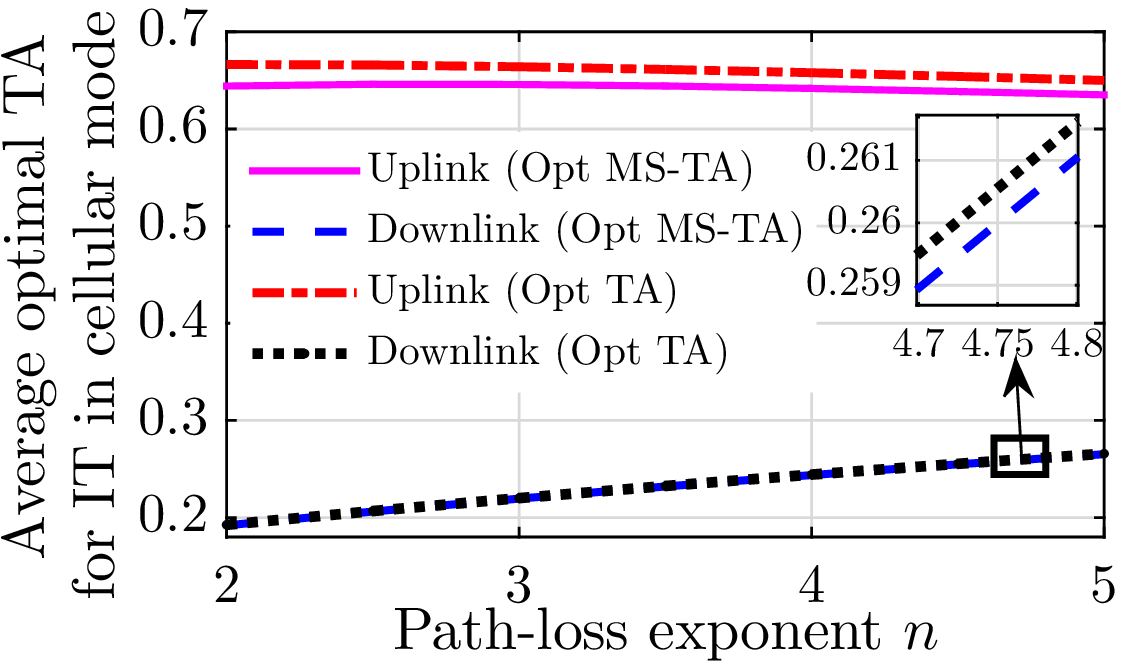} }}
  \subfigure[Approximation validation]{{\includegraphics[width=2.1in]{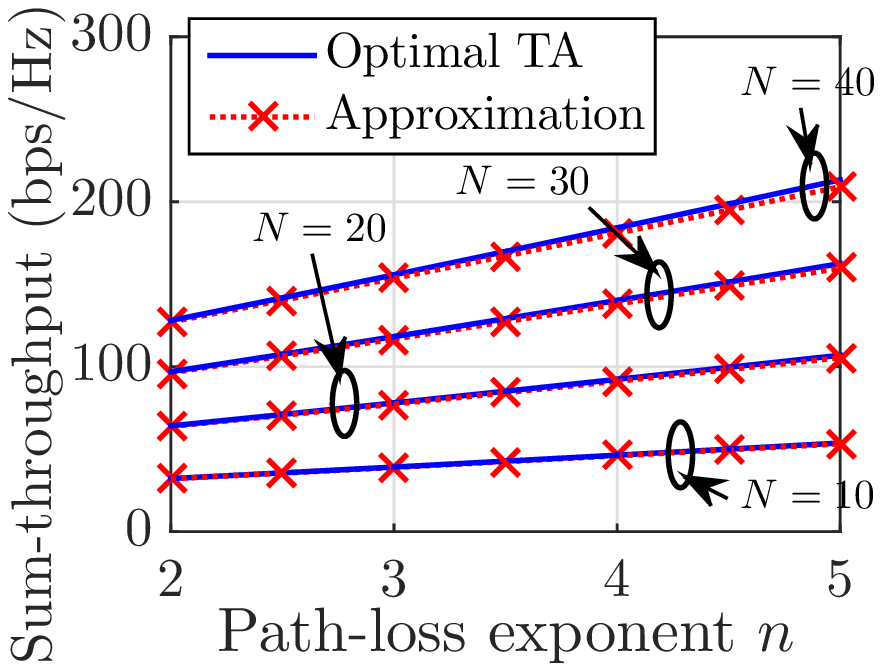} }}
    \caption{\small Insights on average IT in cellular mode and validation of closed-form approximation for optimal TA in all-D2D mode scenario.}
    \label{fig:Opt}
\end{figure}

We numerically quantify the average optimal uplink and downlink IT times in cellular mode. Fig.~\ref{fig:Opt}(a) shows that optimal uplink IT time is much higher than optimal downlink IT time due to significantly low link quality for uplink IT from energy constrained users. On the contrary, downlink involves IT from energy-rich HAP. Further, the IT times (of both uplink and downlink) for joint MS and TA are relatively lower than that for scenario where all nodes follow cellular mode because the former selects only the pairs that have better uplink and downlink qualities for cellular mode communication. 

From Figs.~\ref{fig:sumrate} and \ref{fig:OD2D}, it was noted that optimal TA with all nodes selecting D2D performs very close to jointly optimal MS and TA. So, in Fig.~\ref{fig:Opt}(b) we have compared the variation of $\tau_{\rm{S}}$ achieved by optimal TA scheme with fixed D2D mode for all nodes against that achieved by considering tight analytical approximation $\widehat{t_e^{\rm{D}^*}}$ for $t_e^*$ as obtained by substituting the average channel gain for the RF-powered D2D IT link in \eqref{eq:D2D-TA}. So, $\widehat{t_e^{\rm{D}^*}}\triangleq\left[1-\frac{\mathcal{Y}_4\mathrm{W}_0\left(\frac{\mathcal{Y}_4-1}{\exp(1)}\right)}{\mathrm{W}_0\left(\frac{\mathcal{Y}_4-1}{\exp(1)}\right)-\mathcal{Y}_4+1}\right]^{-1}$ where $\mathcal{Y}_4\triangleq\sum\limits_{i=1}^{N/2}\frac{2\,\mathcal{Y}_{3,i}}{N}$. Fig.~\ref{fig:Opt}(b) shows that although the quality of approximation gets degraded with increasing $N$, it is still very much acceptable as the average percentage error is always less than $1.5\%$. 
 
\begin{figure}[!t]
\centering{{\includegraphics[width=3.3in]{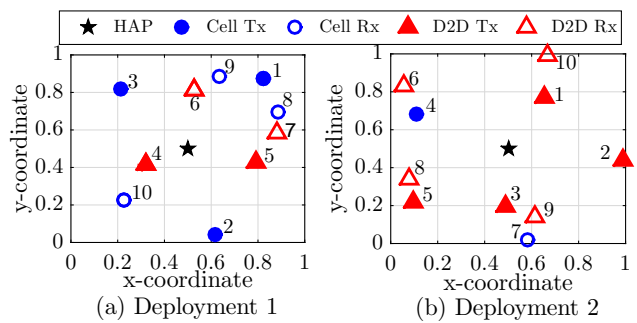} }} 
\caption{\small Examples to give graphical insights on optimal MS strategy.}
    \label{fig:dep} 
\end{figure}
Next we present graphical insights on optimal MS. In Fig.~\ref{fig:dep}  we have plotted two $N=10$ user deployments with $L=1$ unit and shown  optimal MS in each case. User pairing is same as discussed before, i.e., user $1$ transmits to $10$, $4$ transmits to $7$, and so on. We note that when transmitter (Tx) and receiver (Rx) are placed almost opposite to each other with the HAP being in the center and far from the Tx, then cellular mode is preferred. Otherwise, mostly ($\approx 
66\%$ times) D2D is preferred. 

\begin{figure}[!t]
\centering
{{\includegraphics[width=3.45in]{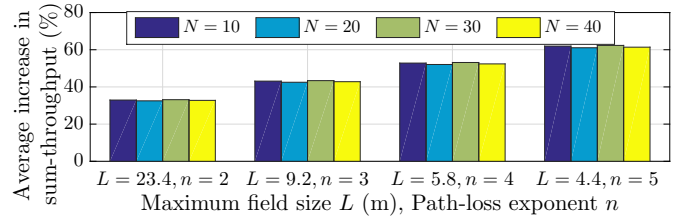} }}
\caption{\small Performance enhancement provided by joint optimal MS-TA.}
    \label{fig:Perf_Impr} 
\end{figure}
Finally via Fig.~\ref{fig:Perf_Impr}, we show that the joint optimization scheme taking advantage of proximity can achieve spectral efficiency gains by optimal MS and can effectively solve the tradeoff between efficient ET and IT by optimal TA. As a result it provides significant gain in terms of $\tau_{\rm{S}}$ over that achieved by benchmark scheme having uniform TA for all three phases (ET, uplink and downlink IT) with all nodes selecting cellular mode. Fig.~\ref{fig:Perf_Impr} also shows that higher gains are achieved when the average link qualities become poorer due to increased $n$.
%\end{eqnarray} 

\section{Conclusion}
We have presented a novel system architecture and transmission protocol for efficient the RF-powered D2D communications. To maximize the sum-throughput of RF-EH small cell OFDMA network, we have derived the joint global optimal MS and TA by resolving the underlying combinatorial issue. Analytical insights on the impact of harvested energy on the optimal decision-making have been provided. We have  observed that the jointly optimal MS and TA can provide about $45\%$ enhancement in achievable sum-throughput. Lastly, we have showed that with our proposed joint MS and TA about $66\%$ nodes follow D2D mode, and the optimal TA scheme with fixed D2D mode for all nodes very closely follows the sum-throughput performance of the jointly optimal scheme.

\section*{Acknowledgments}
This work was supported by the Department of Science and Technology under Grant SB/S3/EECE/0248/2014 along with the 2016 Raman Charpak and 2016-2017 IBM PhD Fellowship programs. In addition, the views of Dr. G. C. Alexandropoulos expressed here are his own and do not represent Huawei's ones.
 
\makeatletter
\renewenvironment{thebibliography}[1]{%
  \@xp\section\@xp*\@xp{\refname}%
  \normalfont\footnotesize\labelsep .5em\relax
  \renewcommand\theenumiv{\arabic{enumiv}}\let\p@enumiv\@empty
  % NEW
  \list{\@biblabel{\theenumiv}}{\settowidth\labelwidth{\@biblabel{#1}}%
    \leftmargin\labelwidth \advance\leftmargin\labelsep
    \usecounter{enumiv}}%
  \sloppy \clubpenalty\@M \widowpenalty\clubpenalty
  \sfcode`\.=\@m
}{%
  \def\@noitemerr{\@latex@warning{Empty `thebibliography' environment}}%
  \endlist
}
\makeatother
\bibliographystyle{IEEEtran}
\bibliography{refs_RFP-D2D_JMSTA}
\end{document}